 \documentclass[12pt,article,superscriptaddress,showkeys]{revtex4-1}
\usepackage{amsmath}
\usepackage{amssymb}
\usepackage{bm}
\usepackage{epsfig}
\usepackage{epstopdf}
\usepackage{graphicx}
\usepackage{color}
\usepackage{mathtools}
\usepackage{float}
\usepackage{placeins}
\usepackage{pifont}
\usepackage{wasysym}
\usepackage{rotating}
\usepackage{gensymb}
\usepackage{multirow}
\usepackage{tabularx}

\usepackage[english]{babel}

\setcitestyle{super}

 \usepackage{color}

\usepackage[normalem]{ulem} 
\usepackage{soul}

\usepackage{calc}
\newsavebox\CBox
\newcommand\hcancel[2][0.5pt]{%
  \ifmmode\sbox\CBox{$#2$}\else\sbox\CBox{#2}\fi%
  \makebox[0pt][l]{\usebox\CBox}%
  \rule[0.5\ht\CBox-#1/2]{\wd\CBox}{#1}} 

\def\be{\begin{equation}}
\def\ee{\end{equation}}
\def\bea{\begin{eqnarray}}
\def\eea{\end{eqnarray}}

\begin{document}
\newcount\timehh  \newcount\timemm
\timehh=\time \divide\timehh by 60
\timemm=\time
\count255=\timehh\multiply\count255 by -60 \advance\timemm by \count255

\title{Circular dichroism in non-chiral metal halide perovskites}

\author{Peter C. Sercel}
\email{psercel@caltech.edu}
\affiliation
{Department of Applied Physics and Materials Science, California Institute of Technology, Pasadena, California 91125, USA}
\affiliation
{ Center for Hybrid Organic Inorganic Semiconductors for Energy, 15013 Denver West Parkway, Golden, CO 80401, USA}
\author{Zeev Valy Vardeny}
\affiliation
{Department of Physics and Astronomy, University of Utah, Salt
Lake City, UT 84112, USA }
\author{Alexander L.  Efros}
\affiliation
{Center for Computational Materials Science, U.~S. Naval Research Laboratory, Washington DC 20375, USA }

\begin{abstract}
We demonstrate theoretically that  non-chiral perovskite layers can exhibit  circular dichroism (CD) in the absence of a magnetic field and without chiral activation by chiral molecules. 
 The effect is shown to be due to splitting of {\it helical } excitonic states  which can  form in structures of orthorhombic or lower symmetry  that exhibit Rashba spin effects. 
The selective  coupling of these helical exciton states  to helical light is shown to give  rise to circular dichroism.
Polarization dependent absorption  is shown to occur due to the combined effect of  Rashba splitting, in-plane symmetry breaking,  and the effect of the exciton momentum  on its fine structure, which takes the form of Zeeman splitting in an effective magnetic field.  
 We calculate significant  CD with an anisotropy factor of up to 30\% in orthorhombic perovskite  layers  under off-normal top   illumination conditions, raising the possibility of  its observation  
 in   non-chiral perovskite structures.   
 
\end{abstract}

\maketitle

An object, a molecule or  a crystal structure is chiral if it is distinguishable from its mirror image; that is, it cannot be superimposed onto the original by any sequence of pure rotations or translations. 
\cite{Kelvin1894} The universally recognized example of  chiral objects are  human hands because the left hand is a non-superimposable mirror image of the right hand, no matter how the two hands are oriented. In nature, chiral molecules and crystal structures exist in two unmixed geometrical forms (two enantiomers) that are mirror images of each other 
 because their low symmetry does not allowed gradual  transformation from one to the other.  

 The growing attention to   chiral materials is  connected  with 
 their potential applications in the area of chiral optoelectronics and spintronics.
  Chirality is associated with the phenomenon of  circular dichroism (CD)\cite{BerovaCD20} since  enantiomer pairs  absorb 
 left and right circularly polarized light differently\cite{MoloneyChemCommun07,MosheNanoLett16,ChenNatCommun20,AhnAmChemSoc20,
 WangAngewChem20},
  and enantiomers can also emit circularly polarized light  in the absence of an external magnetic field.\cite{HanAdvOptMater18,MaACSNano19,ShiAdvMater18}   
Recently chiral materials have been  used in spintronic devices  as a spin filters  due to chiral-induced spin selectivity.\cite{LuSciAdv19,BloomNanoLetter16,SargentNaturePhoton18}

Chiral non-centrosymmetric molecules and crystals  can be categorized  into 65  Sohncke space groups within different Bravais 
lattices and point groups (see for example the review by Long {\it et al.} in Ref.\citenum{SargentNatureReview20}).   Chirality is very common   among natural organic compounds, such as amino acids, sugar molecules,  peptides and DNA.  However, inorganic
materials that show CD  or circular birefringence are relatively rare. The most famous among them, and the first discovered, is quartz. \cite{Kaminsky2000}
Other examples include zinc blende semiconductors under uniaxial stress.\cite{Koopmans1998}

Semiconductor inorganic nanomaterials, such as quantum dots (QDs) and nanoplatelets have been shown however to exhibit CD 
after ``chiral activation'' of their surface by chiral organic molecules.
\cite{MosheNanoLett16,TohghaACSNano13,MosheACSNano11,GeorgievaAdvMater18} An excellent recent review\cite{AhnMaterHoriz17} discusses different induction mechanisms that have been suggested for
achieving this: (i) chiral-organic-molecule-induced crystallization of nanostructures
into a chiral  structure;\cite{MosheAngewChem13} (ii) surface-chiral-organic-molecule-
induced chiral distortion of the surfaces of QDs;\cite{ZhouChemSoc10} and (iii) electronic interactions between chiral organic molecules and QDs.\cite{TohghaACSNano13}  
Chiral  activation of hybrid organic - inorganic
perovskites (HOIPs) by incorporation of chiral organic molecules within the inorganic framework of these perovskites is the most successful realization  of this idea. \cite{DongSmall19} The first demonstration of chirality in HOIPs was in a
1D chiral- perovskite single crystal in 2003,\cite{BillingActaCrystallogr03} followed by 2D chiral- perovskite single crystals in 2006.\cite{BillingCrystEngComm06} Homogeneous chiral  films can be fabricated from perovskites by using various film-coating
methods underscoring the practically of this approach.\cite{AhnMaterHoriz17} 

 Recently, CD has been  observed   in QDs  and nanorods which have {\it not } been  activated by chiral molecules. \cite{MukhinaNanoLatt15,MiltonNanoHor16,LiNatureCommun20} 
This is surprising since observation of  chiro-optic activity in randomly oriented nanoparticles requires chirality.
To explain this observation the authors of Ref.\citenum{MukhinaNanoLatt15} suggested that the QDs have   chiral defects such as  screw dislocations,  which affect the optical properties of small size CdSe/ZnS quantum dots, inducing the CD.  In contrast, in Ref. \citenum{Lifshitz},  elliptical polarization in the photoluminescence of {\it single} CsPbB$_3$ nanocrystals    at zero magnetic field was attributed to inversion symmetry breaking and the Rashba effect.

\begin{figure}[h!]
\centering
\includegraphics[width=14cm]{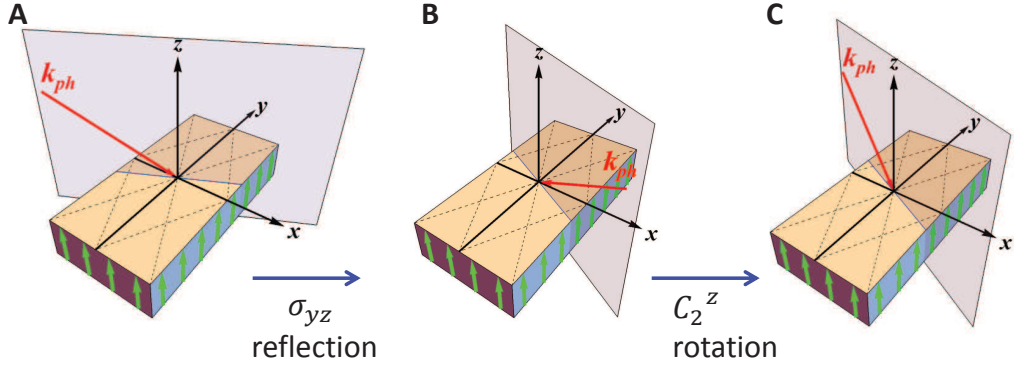}
\caption{ Schematic showing how a non-chiral structure can exhibit chiral behavior when interacting with light.  The figure depicts light (shown via its wave vector $\bm k_{ph}$) incident on a non-chiral perovskite structure  with orthorhombic point symmetry  $C_{2v}$ (depicted as a slab with a two-fold rotational symmetry about the $\hat{z}$ axis).  The structure is assumed to have inversion asymmetry  in the vertical  $\hat{z}$ direction, depicted via  the vertical green arrows in the slab,  while the two vertical mirror planes axes are  $\sigma_{x,z}$ and $\sigma_{yz}$.  In the absence of light, the system is {\it non-chiral }
because of its  mirror symmetries.  On the other hand, the system structure + light, shown in panel A, with light in a plane of incidence   rotated at an arbitrary angle from either of the vertical mirror planes,  is {\it chiral}:   Its mirror image, panel B, cannot be restored to the original by any sequence of rotations or translations. As an example,  panel C shows the effect of two fold rotation about the z axis.
}
\label{Cartoon}
\end{figure}

 It has long been known 
  that symmetry considerations permit 
optical activity and CD 
 in certain non-chiral crystal structures, namely, 
  the four non-chiral crystal classes corresponding to point groups $C_s$, $C_{2v}$, $S_4$, and $D_{2d}$ ($m$, $mm2$, $\bar{4}$, and $\bar{4}2m$ in Hermann- Mauguin notation, respectively).  \cite{Nye,Hobden}
 The first observation  of this phenomenon was  made only in 1967 in  AgGaS$_2$   
  \cite{Hobden} because the optical activity of these  non-chiral crystal structures is weak.
  To understand why   chiro-optic effects
  might   be observed  in  these  non-chiral structures,  let us consider a  crystal  with point symmetry $C_{2v}$.  Although the crystal is non-chiral because it has two mirror planes, if we consider the effect of the measuring light
as shown in Fig. \ref{Cartoon},  we recognize that the  symmetry of the complete light-matter system, without orientational averaging,  is in fact chiral  due to the symmetry breaking associated with the directionality of the light.  
Such light-induced chirality, which has  been termed  ``extrinsic chirality'' \cite{PlumAPL2008}, was recently demonstrated to cause chiro-optic effects in planar metamaterials,\cite{PlumAPL2008,PlumJOSAA2009}    although the planar metamaterial systems 
exhibit reversal of handedness when illuminated from opposite directions.\cite{PlumJOSAA2009,Capasso2018}

In this paper we suggest  a specific physical mechanism that leads to  large CD 
in non-chiral 
 metal halide perovskite crystals 
 with  broken  inversion symmetry, 
  in the absence of any chiral molecules at their surface. 
 Since many perovskites adopt an orthorhombic crystal structure at low temperature, we consider  an orthorhombic system with inversion symmetry breaking normal to the surface  along a two-fold rotation axis:  The structure possesses  non-chiral $C_{2v}$ point symmetry (see Figure \ref{Cartoon}) and is expected to exhibit spin splitting due to the large spin orbit coupling in these systems. 
 Calculations of excitonic  light absorption conducted in  quasi-two dimensional  metal halide perovskites using experimentally determined  
   Rashba spin-orbital splitting parameters \cite{VardenySciAdv,VardenyNatComm2020}
    show  CD  with an anisotropy factor\cite{SargentNatureReview20}  of up to 30\%.
    The effect
     is shown  to be due to  splitting of {\it helical } excitonic states, which  can exist only in structures   
 of orthorhombic (or lower) symmetry with inversion symmetry breaking. 
 The selective  coupling of these helical states  to helical light gives  rise to CD.
 
\newpage

\section{Excitons in  perovskite layers}

Let us consider the problem  of an exciton in a  semiconductor with  parabolic conduction and valence bands.  Introducing the relative electron-hole coordinates  $\bm r = \bm r_e - \bm r_h$ and  the coordinates of the exciton center of mass, $\bm R= (m_e \bm r_e + m_h \bm r_h)/M$, where the exciton translational mass $M=m_e+m_h$ is   equal to the sum of the electron, $m_e$, and  hole, $m_h$ effective masses, respectively, we can describe the exciton energy as a sum of Hamiltonians that describe the  relative electron-hole motion, $\hat{H}_{0,{\rm REL}},$ and  the exciton center of mass (COM) motion $\hat{H}_{0,{\rm COM}}$:
\be
\hat{H}_0 = \hat{H}_{0,{\rm COM}} + \hat{H}_{0,{\rm REL}}={\hat {\bm P}^2\over 2M}+ \left[{\hat {\bm p}^2\over 2\mu}+ V(|\bm r|)\right]~.
\label{eqHamSeparated}
\ee  
Here, $\hat {\bm P}=\hat {\bm p}_e+\hat {\bm p}_h$  is the COM momentum while  $\hat{\bm p} = (m_h\hat{\bm p}_e- m_e \hat{\bm p}_h)/M$ is the relative coordinate momentum, and $\mu=(1/m_e+1/m_h)^{-1}$ is the reduced mass.
As a result of this separation, the  wave function of  a  free exciton  can be written in the form of a Bloch wave.   In the case  of a quasi-2D perovskite,  the wave function takes the form,
\begin{align}
\Psi_{\bm K, n,m; j_e,j_h}(\bm r_e,\bm r_h) = u_{j_e}(\bm r_e) u_{j_h}(\bm r_h) \frac{1}{\sqrt{S}} ~e^{i \bm K \cdot \bm R} \phi_{n,m}(\bm r_e - \bm r_h)~.
\hspace{0.3in}
\label{eqExcPlane2D}
\end{align}
In this expression, wave vector $\bm K$ is equal to $\bm P/\hbar$, $S$ is the surface area of the 2D layer  and  $u_{j_e}$ and  $u_{j_h}$   are the band-edge periodic basis functions for the electron and hole, respectively.    The  indices $j_e$, $j_h$  in $u_{j_e}$ and  $u_{j_h}$ represent in abbreviated fashion the angular momentum quantum numbers $j, j_z$ associated with the electron and hole  Bloch functions.   In Eq. \eqref{eqExcPlane2D} $\phi_{n,m}$ is the normalized wave  function  that describes the relative motion of the electron and hole and is characterized   by principle and azimuthal quantum numbers $n$ and $m$.   The energy spectrum of the 2D exciton consequently can be written as:
\be
E_{n,|m|}^0(\bm K)  =  \mathcal E_{n,|m|} + \frac{  \hbar^2 \bm K^2}{2 M}~, 
\label{eq:4}
\ee
 where $\mathcal E_{n,|m|}$ are  the eigenvalues associated with the internal exciton motion.    In  the absence of dielectric confinement, that is, when the dielectric constants of semiconductor, $\kappa,$ and the surrounding medium are equal,  the  Coulomb potential  can be written as $V(\rho)=e^2/\kappa\rho$, where $\rho= |\bm r_e - \bm r_h|$.  In this limit,  expressions for the energies and wavefunctions of  2D excitons are well known and can be found for example  in Refs. \citenum{EfrosSovPhysSemicond86,YangPRA}.
Due to the small value of the dielectric constant of organic materials, dielectric confinement  can strongly enhance   the electron-hole Coulomb interaction in hybrid organic-inorganic layered perovskite semiconductors.  Analytical expressions for the Coulomb interaction  potential  $V(\rho)$   which account for dielectric confinement effects  were derived by Ritova\cite{Ritova67} and Keldish\cite{Keldish79}. The enhancement of the exciton binding energies for    several low-energy transitions has been demonstrated in CdSe  nanoplatelets \cite{GipiusPRB14}

\section{Rashba effect in a   perovskite layer}
 Let us now consider a  metal halide perovskite  with orthorhombic crystal symmetry and broken  inversion symmetry in  the $z$ direction  perpendicular to the layer. \cite{VardenyNatComm2020}  Due to the strong spin-orbit coupling in metal halide perovskite semiconductors,  the  broken inversion symmetry generates Rashba terms that are   linear in the  electron and hole momenta.      It was shown Ref. \citenum{Becker} that in general, for crystals that have orthorhombic symmetry  with an arbitrary symmetry breaking direction  a total of 12 terms (6 for electron operators, 6 for hole operators) are required for a general description of  the Rashba effect for excitons. In the case that  inversion symmetry is broken in  the $z$ direction the
Rashba Hamiltonian is significantly simplified and is described just by two Rashba coefficients, $\alpha^{e}_{xy}$ and  $\alpha^{e}_{yx}$  for the conduction band and two Rashba coefficients, $\alpha^{h}_{xy}$ and  $\alpha^{h}_{yx}$,  for the valence bands: 
\be
\hat{H}_R = \frac{\alpha^{e}_{xy}}{\hbar} J_x \hat{p}_y^e - \frac{\alpha^{e}_{yx}}{\hbar}  J_y \hat{p}_x^e + \frac{\alpha^{h}_{xy}}{\hbar}  s_x \hat{p}_y^h - \frac{\alpha^{h}_{yx}}{\hbar} s_y \hat{p}_x^h
\label{eqRashbaZ}
\ee
Here, $J_{x,y}$ and $s_{x,y}$ are  the $x$ and $y$  Pauli operators corresponding  to the angular momenta of the electron and hole , respectively,  while $\hat{p}_{y,x}^{e,h}$ are the  projections of the electron and hole momentum  operators on the $x$ and $y$ axes. For analysis of the effect of Rashba terms on the free exciton, it is useful to express the  electron and hole momentum operators in terms of  COM and relative momenta.  Using
$\hat{\bm p}_e  = (m_e/M) \bm P + \hat{\bm p}$ and
$\hat{\bm p}_h = (m_h/M) \bm P -\hat{\bm p},$
we  transform the Rashba Hamiltonian in Eq. \eqref{eqRashbaZ} into a sum of Rashba Hamiltonians for  the exciton center of mass motion, $\hat{H}_{R,{\rm COM}}$, and the  electron -hole relative motion,  $\hat{H}_{R,{\rm REL}}$, where
 \bea 
 \hat{H}_{R,{\rm COM}}(\bm P)&= &{1\over M}\left[\left(m_e \frac{\alpha^{e}_{xy}}{\hbar}  J_x +  m_h\frac{\alpha^{h}_{xy}}{\hbar}  s_x\right) P_y -
\left(m_e \frac{\alpha^{e}_{yx}}{\hbar}  J_y +  m_h\frac{\alpha^{h}_{yx}}{\hbar} s_y\right) P_x\right]  \nonumber \\
 \hat{H}_{R,{\rm REL}}&= &\left(\frac{\alpha^{e}_{xy}}{\hbar}  J_x -\frac{\alpha^{h}_{xy}}{\hbar}  s_x\right) \hat{p}_y -
\left(\frac{\alpha^{e}_{yx}}{\hbar}  J_y - \frac{\alpha^{h}_{yx}}{\hbar} s_y\right) \hat{p}_x\hspace{1.1in}
\label{eqRashbaZSep}
\eea
The effect of the Rashba Hamiltonian  $\hat{H}_{R,{\rm REL}}$ on the fine structure of the  exciton due to internal electron-hole motion  can be described in the framework  developed for the 3D exciton in Ref. \citenum{Becker}.  There, it was demonstrated that when  Rashba terms exist in both the conduction and valence bands,  an effective  $\bm J \cdot \bm s$ interaction can invert the level order of the bright and dark exciton fine structure levels.

 From now on however, for simplicity,  we will consider  Rashba terms only in the conduction band, neglecting the effect of spin orbit coupling in the valence band,  assuming that $\alpha_{xy}^h\approx 0$ and $\alpha_{yx}^h\approx 0$. This is reasonable given that the valence band states transform with overall $s$ symmetry and are largely comprised of the Pb $6s$ atomic orbitals. \cite{Becker}   In this approximation, the Rashba terms  $\hat{H}_{R,{\rm REL}}$   originating from the motion associated with the relative electron-hole coordinate  do  not affect  the  exciton fine  structure.\cite{Becker}   Only the  Rashba terms associated with COM motion of the exciton enter into the present analysis.  These terms, along with electron-hole exchange,   determine the exciton fine structure.  
 One can see  from Eq.\eqref{eqRashbaZSep} that  the Rashba Hamiltonian for the exciton center of mass motion, $\hat{H}_{R,{\rm COM}}$,  does not mix states with different momenta,  $\bm P = \hbar \bm K$, but it does mix  the four angular momentum sublevels of the exciton fine structure created by the electron-hole exchange interaction.  In this limit Rashba Hamiltonian $\hat{H}_{R,{\rm COM}}$ in Eq. \eqref{eqRashbaZSep} reduces to,
 \begin{align}
 \hat{H}_{R,{\rm COM}}(\bm K)= \frac{ m_e}{M} \left( \alpha_{xy} K_{y}  J_x  - \alpha_{yx} K_{x} J_y\right)~,
 \label{RashbaGenZ2D}
 \end{align}
where we have now dropped the superscript $e$ on the Rashba coefficients with the understanding that henceforth Rashba coefficients refer to the conduction band. Equation \eqref{RashbaGenZ2D}  can be rewritten as sum of 2D Rashba $\hat{H}_{R}^{2D}(\bm K)$ and 2D Dresselhaus $\hat{H}_{D}^{2D}(\bm K)$  terms 
that are familiar from studies of the effects of structural inversion asymmetry in 2D electron systems: 
\cite{RashbaPRL03,Ganichev14}
 \begin{align}
 \hat{H}_{R,{\rm COM}}(\bm K)= \hat{H}_{R}^{2D}(\bm K)+\hat{H}_{D}^{2D}(\bm K)~,
 \end{align}
where
 \begin{align}
 \hat{H}_{R}^{2D}(\bm K)= {m_e \alpha\over M} (K_{y} J_x  - K_{x}J_y) \hspace{0.4in} {\rm (Pure\hspace{0.05in} 2D-Rashba);}\\
 \hat{H}_{D}^{2D}(\bm K)= \frac{ m_e \beta}{M} (J_x  K_{y}  + J_y K_{x} )\hspace{0.3in} {\rm (Pure\hspace{0.05in} 2D-Dresselhaus)}~,
 \end{align}
  with $\alpha = (\alpha_{xy}+\alpha_{yx})/2$ and  $\beta = (\alpha_{xy}-\alpha_{yx})/2$.  
In a system with cubic or tetragonal symmetry, $\alpha_{yx}=\alpha_{xy}=\alpha$ since these systems have  four-fold rotational symmetry about the z-axis;  the $x$ and $y$ directions are equivalent. We call this the ``pure'' Rashba limit.
   However,  if the $x$ and $y$ directions are inequivalent, as is the case   with orthorhombic  $C_{2v}$ point symmetry, then in general, the two Rashba coefficients may be distinct: $\alpha_{yx} \neq\alpha_{xy}$.  In the extreme limit that $ \alpha_{xy} = - \alpha_{yx} =\beta$ the Rashba part of Hamiltonian \eqref{RashbaGenZ2D} vanishes, $\hat{H}_{R}^{2D}(\bm K)\equiv 0$, and we have the  pure 2D Dresselhaus Hamiltonian $\hat{H}_{D}^{2D}(\bm K)$.

To gain physical intuition on the splitting of the exciton levels,  it is convenient to introduce an  effective  $\bm K$ dependent magnetic field $\bm B^{eff}(\bm K)$ which allows us  to write the Rashba Hamiltonian in Eq.  \ref{RashbaGenZ2D} in the form of the Zeeman effect, $\hat{H}_{R,{\rm COM}}= g^{eff} \mu_B \bm B^{eff}\cdot \bm J,$ 
 where $\mu_B$ is the Bohr magneton, $g_{eff}=m_e/M$~, and the effective magnetic field is given by,
\begin{align}
\bm B_{eff}(\bm K) \equiv  \frac{1}{\mu_B}\left(\alpha_{xy} K_{y} \hat{x} - \alpha_{yx} K_{x} \hat{y} \right)~.
\label{eqEffBfield}
\end{align}

One can see that  the orientation of this magnetic field  depends strongly on $\bm K$.     The magnitude of the effective magnetic field $B_{eff}= \sqrt{\alpha_{xy}^2K_y^2+\alpha_{yx}^2K_x^2}/\mu_B$  may take on quite significant values.
 We estimate it using the conduction band Rashba energy of $40$meV measured by Zhai et al.  in the 2D layered perovskite $(\rm PEA)_2PbI_4$ (PEPI) and assume the electron and hole effective masses to be $0.25 m_0$. \cite{VardenySciAdv}  With light of free space wavelength $\lambda_0 = 500$nm incident at $45$ degrees from vertical, assuming a refractive index $n=\sqrt{5}$, corresponding to a high frequency dielectric constant of $\epsilon_\infty = 5$, which is typical of metal halide perovskites, the in-plane wave vector of excitons created by absorption is $K \sim 0.009 nm^{-1}$.  This leads to an effective magnetic field of 24  Tesla.
The resulting splitting of the exciton levels can be described  using a basis of total angular momentum, $\bm F$, and its projection $F_{z_B}$, where the  quantization axis $\bm z_B$ is  the direction of the effective magnetic field: $\bm z_B=\bm B_{eff}/B_{eff}$.
 
\begin{figure}[h!]
\centering
\includegraphics[width=12cm]{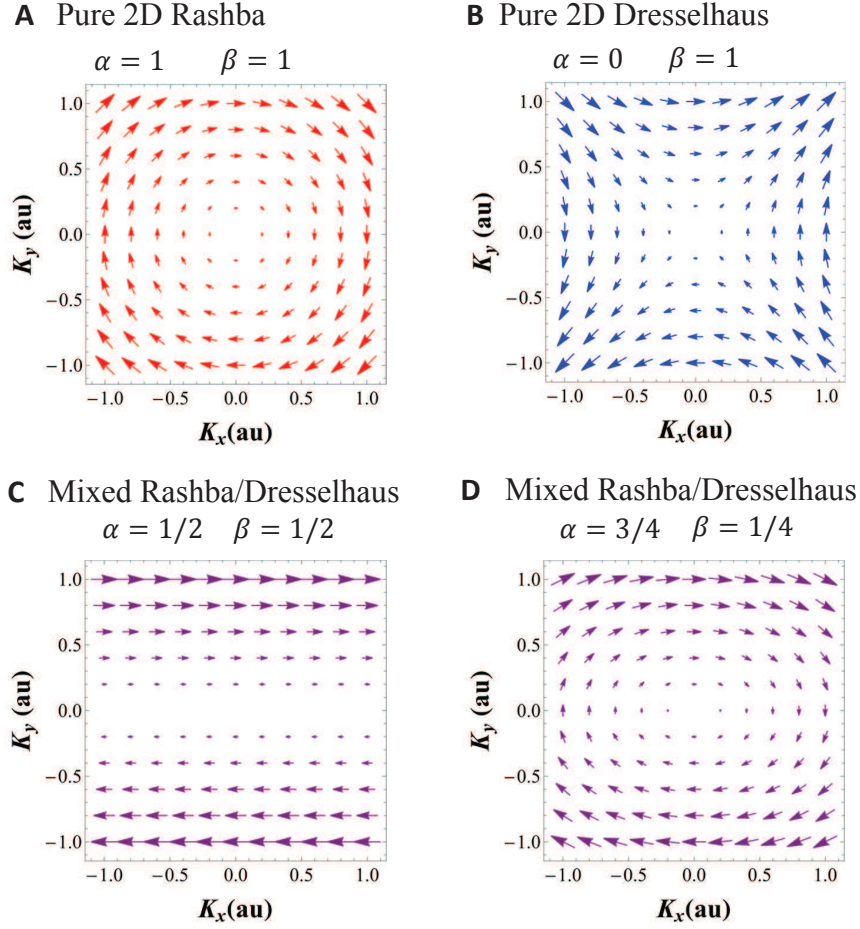}
\caption{
 Vector field maps of the effective magnetic field due to exciton center of mass motion in the presence of Rashba splitting.   The figure shows the relative magnitude and direction of the effective magnetic field defined in Eq. \ref{eqEffBfield} in the $K_x$, $K_y$ plane.  The $K_x$ and $K_y$ axes are plotted in arbitrary units (au) in the figure. Panel  A  shows the relative effective magnetic field map for  the pure Rashba case,  ($\alpha=1$, $\beta =0$ in dimensionless units), while panel B shows the pure 2D Dresselhaus case ($\alpha  =0$, $\beta =1$).  Panels C and D show mixed cases corresponding respectively to  dimensionless $\alpha  =\beta =1/2$, and $\alpha= 3/4,$, $ \beta =1/4$.   For each case the components of the Rashba tensor $\alpha_{xy}$ and $\alpha_{yx}$  corresponding to  Eq. \ref{RashbaGenZ2D}  are also given.
}
\label{FigBfieldMap}
\end{figure}

Figure \ref{FigBfieldMap}   shows vector field maps depicting the direction and relative magnitude of the effective magnetic field for the pure Rashba  ($\alpha \neq 0$, $\beta =0$) and the pure 2D Dresselhaus ($\alpha  =0$, $\beta \neq0$) cases, as well as some mixed cases.   From Eq. \ref{eqEffBfield} and these field maps,  we can recognize that in the pure Rashba case, the effective magnetic field is always perpendicular to the direction of exciton momentum $\bm K$.  In the pure 2D Dresselhaus case by contrast, the effective magnetic field generally has a component along $\bm K$; on   the lines $K_x = \pm K_y $, $\bm B_{eff}$ is co-linear with the direction of the vector $\bm K$.    As a result, in this case,   triplet excitons with angular momentum projection  $F_{z_B}= \pm 1$   have {\it helicity},
a property which is  defined as $\chi 
= \bm F \cdot \hat{\bm n}_{\bm K}$ where $\hat{\bm n}_{\bm K}=\bm K/K$.   Excitons with angular momentum   parallel  ($F_{z_B}= + 1$) or antiparallel  ($F_{z_B}= - 1$) to the direction of the momentum  $\bm K$ have helicity $\chi =+1$ and $\chi =-1$ respectively. Moreover, under the action of the effective magnetic field these states of opposite helicity are energetically  split. 
The non-zero  helicity and the effective  Zeeman splitting of the $F_{z_B}= \pm 1$ states in an orthorhombic  system gives rise to CD.

 We proceed now to a quantitative analysis of the effect of the Rashba terms on the exciton fine structure and dichroic optical properties. The exciton Rashba Hamiltonian described by  Eq. \ref{RashbaGenZ2D} for non-zero $\bm K$ splits the degeneracy of the 4 exciton states. Taking into account the exciton kinetic energy $\hbar^2K^2/2M$, where $K = \sqrt{K_x^2 + K_y^2}$, the energy  for any momentum direction is determined as a function of $K_x, K_y$ by (see the Supporting Information),
\begin{align}
E_{COM}(K) =  \frac{\hbar^2 K^2}{2 M}\pm\frac{ m_{e}}{M} \sqrt{\alpha_{yx}^2 K_{x}^2+\alpha_{xy}^2 K_{y}^2}~. 
\label{eqCOMrashbaE}
\end{align}
The  COM motion of free excitons described by Eq. \eqref{eqCOMrashbaE}  has the   offset parabolic dispersion,    well known   for 2D electrons in the presence of Rashba terms. \cite{Ganichev14}   

A complete description  of the exciton dispersion requires, however,  that we also include the exciton fine structure splitting due to electron-hole exchange, resulting in  the following  Hamiltonian:
\begin{align}
\hat{H}_{\rm tot}  = \hat{H}_{\rm INT}  + \hat{H}_{R,{\rm COM}}^{D,X,Y,Z}(\bm K)
\label{eqHtot}
\end{align}
where $\hat{H}_{\rm INT}$ describes the fine structure of the exciton connected with exciton internal motion at $K=0$. It  is straightforward  to write $\hat{H}_{\rm tot}$   in a basis of  the four electron and hole  Bloch function products, $|u_{j_e=1/2}\rangle |u_{j_h=1/2}\rangle$, $|u_{j_e=1/2}\rangle |u_{j_h=-1/2}\rangle$,  $|u_{j_e=-1/2}\rangle |u_{j_h=1/2}\rangle$ and $|u_{j_e=-1/2}\rangle |u_{j_h=-1/2}\rangle$.   To study the exciton  polarization properties  it is most  convenient  to  transform into a basis of exciton states whose dipoles are oriented along the $x, y, z$ directions. \cite{Sercel2019jcp} 
This $\cal O$ basis is described by the  $|X\rangle$,  $|Y\rangle$,  and $|Z\rangle$ wave functions of the three exciton states $X$, $Y$, and $Z$ whose dipoles are oriented along the $x$, $y$, and $z$ directions, and the $|D\rangle$ wave function of the dark $D$ exciton state, which is dipole inactive.\cite{Sercel2019jcp}    The unitary transformation to this basis, developed in the Supporting Information section, results  in the following matrix representation,
\bea
&\hat{H}^{\rm tot}_{D,X,Y,Z}(\bm K)& = \left[\mathcal E_{0,0}+\frac{\hbar^2(K_x^2+K_y^2}{2 M}\right] \mathbb{I}+ \nonumber\\
&&\left(
\begin{array}{cccc}
 E_d & -\alpha_{xy} K_{y}  & \alpha_{yx} K_{x}  & 0 \\
 -\alpha_{xy}K_{y}  & E_x & 0 & -i \alpha_{yx} K_{x}  \\
 \alpha_{yx} K_{x}  & 0 & E_y & -i \alpha_{xy} K_{y}  \\
 0 & i \alpha_{yx} K_{x}  & i \alpha_{xy} K_{y}  &E_z \\
\end{array}
\right),~~~~~~
\label{eqRashbaPlusEx}
\eea
where $\mathbb{I}$ is the 4x4 unit matrix.  The second term of Hamiltonian  $\hat{H}^{\rm tot}_{D,X,Y,Z}(\bm K)$ describes the exciton fine structure, where  $E_d$ is   energy of the dark exciton at $\bm K = 0$, while the bright triplet states $X$, $Y,$ and $Z$   have      $\bm K = 0$ energies  $E_x$, $E_y$, and $E_z$ respectively.  These are degenerate in the case of cubic symmetry but split in general in crystals with orthorhombic crystal structure.\cite{Becker,Sercel2019jcp}  In our calculations, for simplicity, we will start by neglecting the crystal field splitting between the $X$, $Y$ and $Z$ exciton states, setting $E_x =E_y=E_z= E_t$, the bright triplet energy.
Although the off-diagonal terms in  Hamiltonian described  by Eq. \eqref{eqRashbaPlusEx} generally  are  not rotationally invariant, we have  found    closed form solutions for the eigenvalues  of  $\hat{H}^{\rm tot}_{D,X,Y,Z}(\bm K)$  in   several special cases.
 In the case that crystal field splitting can be neglected, $E_x =E_y=E_z= E_t$, assuming that  $\alpha_{yx} = \alpha_{xy}$   (the pure Rashba case $\alpha_{xy} = \alpha$), or $\alpha_{yx} = - \alpha_{xy}$ (the  pure Dresselhaus case, $\alpha_{xy} = \beta$)   we obtain,
\bea
E_{1,\pm 1}(\bm K) &=&  \mathcal E_{0,0}+\Delta  +\frac{\hbar^2K^2}{2 M}\pm K {m_e \alpha_{xy} \over M}~,\nonumber\\
E_{1/2\pm1/2,0}(\bm K) &= & \mathcal E_{0,0}+\frac{\hbar^2 K^2}{2 M}+{\Delta\over 2} \pm  {\sqrt{\Delta^2M^2 + 4K^2  (m_e \alpha_{xy})^2}\over2 M}~.
\label{LifshitzTrpure}
\eea
Here,  $\Delta \equiv E_t-E_d$ is the energy difference between the bright triplet states and the dark exciton state.  The subscripts on the energy in this expression identify the  projection of the exciton angular momentum, $F_{z_B}$ along the  quantization axis $\bm z_B$ taken in the  direction of the effective magnetic field $\bm B^{eff}(\bm K)$. The direction $\bm z_B$ thus varies with  $\bm K$  as described  in Eq. \ref{eqEffBfield} and in Fig. \ref{FigBfieldMap}. The analytical  form for the exciton dispersion  can also  be found  in the case that the triplet degeneracy at $\bm K = 0$ is lifted, with $E_z \neq E_t$ (see SI). Generally, however,  for any symmetry, the fine structure of the ground exciton state can be written as $E_j(\bm K)=E_{1,0}^0(K)+\delta_j(\bm K)$, where $E_{1,0}^0(K)$ from Eq. \ref{eq:4} is the exciton energy in the absence of fine structure splitting  $\delta_j(\bm K)$.  The latter is found by diagonalization of   Hamiltonian $\hat{H}_{\rm tot}$ from Eq. \eqref{eqHtot} (for details see SI). 
\be
\Psi_{\bm K, j}(\bm R,\bm r) ={\exp(i \bm K \cdot \bm R)\over \sqrt{S}}~ \phi_{1,0}(\bm r)\sum_{i=D,X,Y,Z}C_{j_e,j_h}^j(\bm K) |u_{i}  \rangle
\label{ExcitonWF}
\ee
where the coefficients $C_{i}^j(\bm K)$ describe the  eigenstate of the corresponding matrix in the  basis of exciton states $|D\rangle$,  $|X\rangle$,  $|Y\rangle$,  $|Z\rangle$.

\begin{figure}[h!]
\centering
\includegraphics[width=13cm]{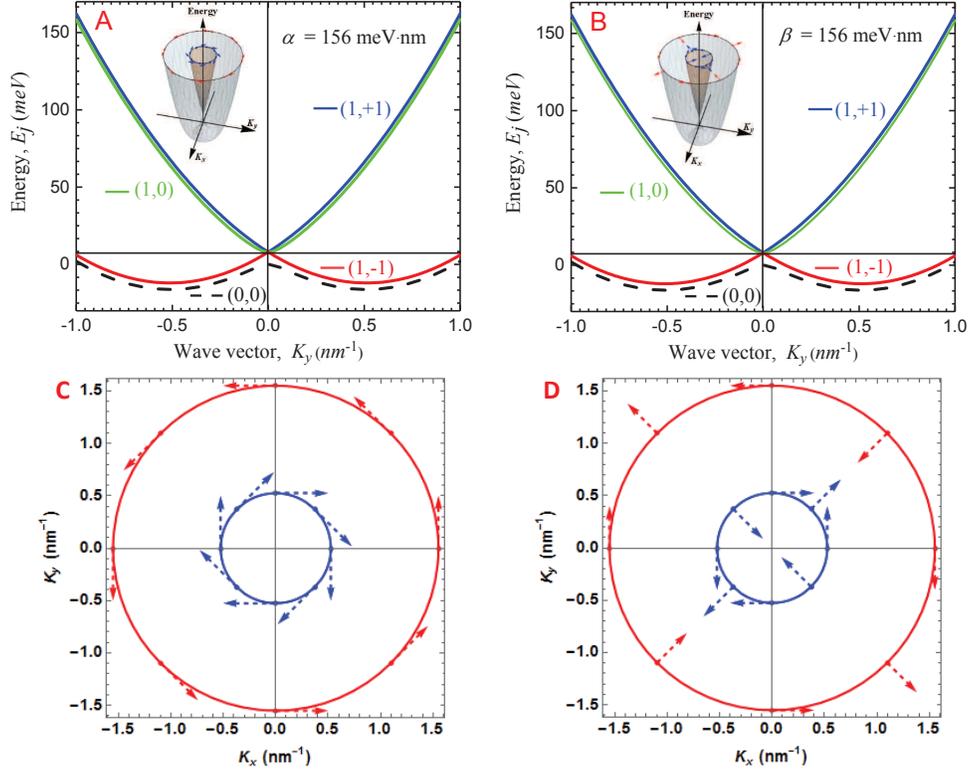}
\caption{Exciton  energies and angular momentum textures in 2D layered  perovskite  with inversion symmetry breaking along the $\hat{z}$ direction normal to the 2D plane.    Panels A and B show the energies calculated  versus wave vector $\bm K = K_y \hat{\bm y}$ for  pure Rashba  and pure 2D-Dresselhaus angular  momentum   textures, respectively, with $\alpha = 156~{\rm meV}\cdot{\rm nm}$ and  $\beta = 156~{\rm meV}\cdot{\rm nm}$  with electron and hole effective masses  $m_e=m_h = 0.25 m_0$, and assuming degenerate  triplet levels at $\bm K=0$ at energy $\Delta =8$ meV
 above the dark state at energy $E_D=0$\,meV.\cite{Tanaka2005} The dispersion curves for each fine structure level  in panels A and B are labelled according to $(F,F_{z_B})$, the exciton  total  angular momentum and  its projection on an axis aligned to the effective magnetic field, see Eq.\ref{eqEffBfield} and Figure \ref{FigBfieldMap}.  The  curves in A and B are identical reflecting symmetry under rotation about the $\hat{z}$ axis  as  evident in the insets of each figure, which show  3D plots of the energy surfaces in the $K_x,K_y$ plane for the levels $(1,\pm1)$.  Panels C and D show the angular momentum textures for these two states along   constant energy contours  at $E=70$ meV. 
}
\label{FigTrousers1}
\end{figure}

Figure \ref{FigTrousers1} shows the exciton  energies calculated   along the  $K_y$ direction using Eq. \ref{LifshitzTrpure}. 
Panels A and B show the energies calculated for  pure 2D-Rashba and pure 2D-Dresselhaus spin textures,  with $\alpha = 156~{\rm meV}\cdot{\rm nm}$ and  $\beta = 156~{\rm meV}\cdot{\rm nm}$, respectively, for electron and hole effective masses   $m_e=m_h = 0.25 m_0$, where $m_0$ is the free electron mass.  These parameters each correspond to a Rashba energy of 40 meV as measured in the 2D HOIS, PEPI.\cite{VardenySciAdv} 
 The calculations shown assume degenerate  triplet levels at $\bm K=0$ at energy $\Delta =8$ meV above the dark state at energy $E_D=0$ meV, corresponding to the   exchange constant $w=12$ meV measured in the 2D hybrid organic
 $\rm (C_6H_{13}NH_{3})_2PbI_4$.\cite{Tanaka2005}  The dispersion curves for each level  are labelled according to $F,F_{z_B}$, the exciton total  angular momentum and  its projection on an axis aligned to the $\bm K$ dependent effective magnetic field $\bm B^{eff}(\bm K)$, see Eq. \ref{eqEffBfield} and Figure  \ref{FigBfieldMap} which shows the field direction in the $K_x$, $K_y$ plane. 
 Referring to Fig. \ref{FigBfieldMap}, we call  attention in particular to the fact that the direction of $B^{eff}$reverses in passing through the origin; with our state labelling convention, the $F_{z_B}=+1$ state always has higher energy than the one with $F_{z_B}=-1$. The dispersion in A and B are identical reflecting   isotropy about the $\hat{z}$ axis.  This symmetry  is  evident in the insets of each figure, which show  3D plots of the energy surfaces in the $K_x,K_y$ plane for the levels $(F,F_z)=(1,\pm1)$.  Panels C and D show the angular momentum textures for these two states along   constant energy contours plotted  at $E=70$ meV, chosen simply for clarity of the display. The direction of the effective magnetic field depends differently on $K_x, K_y$ for the pure Rashba versus the pure Dresselhaus spin textures (see Figure  \ref{FigBfieldMap}), resulting in the distinctly different exciton angular momentum textures shown in panels C and D. The $F_{z_B}=\pm 1$ exciton states for the pure Dresselhaus case, panel D, exhibit {\it helicity} for any wave vector   not aligned to the mirror plane directions $\hat{x}$ and $\hat{y}$, with maximum   helicity realized along the lines $K_x=\pm K_y$.  The corresponding exciton states for the pure Rashba case are achiral since the angular momentum is orthogonal to $\bm K$ for all $\bm K$. 

The significance of the helicity consideration is underscored by examination of the exciton eigenvectors, transition dipoles and helicity corresponding to Fig.\ref{FigTrousers1}.  Table  \ref{tab:1}  shows these  for the case  $E_z = E_t$ for  both the Rashba  angular momentum  texture and the pure 2D Dresselhaus  angular momentum texture, respectively. The energy eigenvalues are  given by Eq. \ref{LifshitzTrpure}  along the line $K_y=K_x$ for positive $K_x$, using $\alpha = \alpha_{xy}=\alpha_{yx}$ for the Rashba case, and $\beta = \alpha_{xy}=-\alpha_{yx}$ for the Dresselhaus case.  The table  shows the  expansion coefficients of the exciton wavefunction in the  $|D\rangle$,  $|X^\prime\rangle$,  $|Y^\prime\rangle$,  and $|Z\rangle$ basis, which is rotated so that the $x^\prime$ axis is  aligned with $\bm K$.  The polarization properties of  each exciton state  is shown in the last column, also given in the rotated coordinate system. While there are two  transition dipoles in the Rashba case  (for the $F_{z_B}=\pm 1$ states)  that {\it appear} circular ($\hat{x}^\prime \pm i \hat{z}$),  they are not actually helical because  the components of the transition dipoles parallel to the $x^\prime$ direction cannot   couple  to transverse  electro-magnetic  waves propagating in the  $x^\prime$ direction. 
These results actually hold for the Rashba case for any wave vector $\bm K$ due to the rotational invariance of Eq. \ref {RashbaGenZ2D} when $ \alpha_{xy}=\alpha_{yx}$.
For the pure
Dresselhaus angular momentum texture  however, it is  the $Y^\prime $ and $Z$ states that are mixed with each other   with a relative phase factor of $i$.  Since these dipole components are both normal to the direction of propagation $\bm K$ along the $x^\prime$ direction,  these states have  circularly polarized transition dipoles. Since they are split, these transitions  exhibit CD as discussed above in the context of the effective magnetic field description. 

This consideration  shows that  the magnitude of  CD    depends on the exciton propagation angle $\phi$ relative to the mirror symmetry planes. Indeed, if this system were to be side illuminated along the $\phi=45^\circ$ azimuth the $F_{Z_B}=\pm1$ states would exhibit 100\% degree of circular polarization. In  Fig. \ref{Fig4CD} we show the angular dependence  of the polarization structure of the exciton sublevels calculated taking into account both Rashba and Dresselhaus terms as well as the exciton fine structure connected with exciton internal motion. 
\begin{table}[]
\begin{tabular}{|c|c|c|c|}
\hline 
\multicolumn{4}{|c|}{Pure 2D Rashba, $\alpha=\alpha_{xy}=\alpha_{yx}$ } \\ 
\hline 
Energies & \{$D$,~$X^\prime$,~$Y^\prime$,~$Z$\} & Polarization, $\bm P_j/P_j$&Helicity $\chi$\\
\hline 
$E_{1,+1}(\bm K)$ & $\frac{1}{\sqrt{2}}\{0,1,0,+i \} $
& $\frac{1}{\sqrt{2}} (\hat{x}^\prime-i \hat{z})$ &0 \\
$E_{1,-1}(\bm K)$ & $\frac{1}{\sqrt{2}}\{0,1,0,-i \} $
& $\frac{1}{\sqrt{2}} (\hat{x}^\prime+i \hat{z})$ &0\\
$E_{1,0}(\bm K)$ &
$ \frac{1}{\sqrt{N_{1,0}}}\{-\frac{\Delta M-\sqrt{\Delta^2M^2 + 4K^2 (m_e \alpha)^2}}{2K m_e \alpha},0,1,0\}$& $ \hat{y}^\prime$ &0 \\ 
$E_{0,0}(\bm K)$ &
$ \frac{1}{\sqrt{N_{0,0}}}\{\frac{\Delta M-\sqrt{\Delta^2M^2 + 4K^2 (m_e \alpha)^2}}{2K m_e \alpha},0,1,0\}$& $ \hat{y}^\prime$ &0\\ 
\hline 
\multicolumn{4}{|c|}{Pure 2D Dresselhaus, $\beta= \alpha_{xy}=- \alpha_{yx}$ } \\ 
\hline 
$E_{1,+1}(\bm K)$ & $\frac{1}{\sqrt{2}}\{0,0,1,+i \} $
& $\frac{1}{\sqrt{2}}(\hat{y}^\prime-i \hat{z})$ &+1 \\
$E_{1,-1}(\bm K)$ & $\frac{1}{\sqrt{2}}\{0,0,1,-i \} $
& $\frac{1}{\sqrt{2}}(\hat{y}^\prime+i \hat{z})$ &-1\\
$E_{1,0}(\bm K)$ &$ \frac{1}{\sqrt{N_{1,0}}}\{\frac{\Delta M-\sqrt{\Delta^2M^2 + 4K^2 (m_e \beta)^2}}{2K m_e \beta},1,0,0\}$& $ x^\prime$ &0\\ 
$E_{0,0}(\bm K)$ &$ \frac{1}{\sqrt{N_{0,0}}} \{-\frac{\Delta M-\sqrt{\Delta^2M^2 + 4K^2 (m_e \beta)^2}}{2K m_e \beta},1,0,0\}$& $ x^\prime$ &0 \\ \hline
\end{tabular}
\caption{Energies, wave functions, polarizations and helicity of exciton states with wave vector $\bm K$ along the line $K_y=K_x$ for positive $K_x$, for the pure 2D Rashba spin-texture, top, and the pure 2D Dresselhaus spin texture, bottom. Energies labelled according to the total angular momentum $F$ and its projection along the effective magnetic field direction, $F_{z_B}$, and are given for each state by Eq. \ref{LifshitzTrpure}. For the Rashba case, $\alpha=\alpha_{xy}=\alpha_{yx}$, while for the Dresselhaus case, $\beta= \alpha_{xy}=- \alpha_{yx}$ in Eq. \ref{eqCOMrashbaE}. Results are shown neglecting the crystal field splitting between the $Z$ exciton and the $X$ and $Y$ excitons ($E_z=E_t$). The table shows the energy and the expansion coefficients of the exciton wavefunction in the $|D\rangle$, $|X^\prime\rangle$, $|Y^\prime\rangle$, and $|Z\rangle$ basis (see Eq.\ref{ExcitonWF}), which is rotated so that the $x^\prime$ axis is aligned with $\bm K$. The factors $N_{1,0}$ and $N_{0,0}$ are normalization coefficients. The polarization properties of each exciton state are given in the rotated coordinate system. The last column gives the helicity, $\chi 
= \bm F \cdot \hat{\bm n}_{\bm K},$ for each state. }
\label{tab:1}
\end{table}

 \section{Calculation of   circular dichroism}
 The analysis conducted in the previous section showed that both cubic  and tetragonal symmetry  are  too high   to observe circular dichroism connected  with exciton COM motion at zero magnetic field.  Intrinsic circular dichroism in perovskites   requires   orthorhombic symmetry, which is required to realize 2D Dresselhaus angular momentum textures. To describe the CD, let us consider the probability of excitation  of 2D excitons by absorption of  light.  We   consider the ground  state of the exciton  in the $x,y$ plane of the 2D layer created by absorption of light  incident   on the 2D layer with wave vector $\bm k_{ph}$  at angle $\theta$ measured from the vertical direction $\hat{\bm n}_z$ (Fig \ref{Cartoon}).  

The  light interacting  with the 2D perovskite layer,  may be absorbed to create an exciton.   The probability of  excitation   of  the $j^{\rm th}$ exciton sub-level of  the ground 2D exciton state  with   in-plane wave vector $\bm K$    can be described by Fermi's golden rule:
\be
{\cal W}_{\bm K, j}={2\pi\over \hbar}|\langle \Psi_{\bm K, j} |\hat{H}_{\rm int} |G \rangle|^2\delta(E_j(\bm K)-\hbar\omega)~.   
\label{eqFGR2d} 
\ee
Here,  $|G\rangle$ is the crystal ground state,
$\hbar \omega$ is the energy of  the absorbed photon, and $E_j(\bm K)$ is the energy of the $j^{\rm th}$ exciton sub-level.
The light-matter  interaction Hamiltonian $\hat{H}_{\rm int}= -(e/m_0 c) \bm A^m \cdot \hat{\bm p}$ is expressed as usual in terms of the inner product of the dipole operator $\hat{\bm p}$ and the vector potential $\bm A^m$ inside the 2D layer. 
 Evaluating the matrix element in Eq. \ref{eqFGR2d} for the exciton state $\Psi_{\bm K,j}(\bm r_e,\bm r_h)$ from Eq.\eqref{ExcitonWF} we find,
 \begin{align}
\langle \Psi_{\bm K,j}|\hat{H}_{\rm int} |G \rangle =   \frac{e}{m_0c} \bm A^m\cdot \bm P_{j}(\bm K) \phi_{1,0}(0) \delta_{\bm k_{ph,\parallel},\bm K}
\label{eqOME2dfinal}
\end{align}
The Kronecker delta represents the well-known momentum conservation  rule, reflecting conservation of momentum in the 2D layer.  The  wave vector  $\bm K$  of the in-plane  exciton created by absorption of a photon must match the in-plane component $\bm k_{ph,\parallel}= \bm k_{ph} -\hat{\bm n}_z (\bm k_{ph}\cdot \hat{\bm n}_z),$ of the photon that was absorbed, and has magntitude $k_{ph} \sin\theta$.
 The contact term  $ \phi_{n,0}(0)\neq 0$ in Eq. \ref{eqOME2dfinal} only for states with internal motion azimuthal quantum number $m=0$ (``s''states).

The key factor in Eq \ref{eqOME2dfinal} that is responsible  for the polarization properties  of absorption  is the dipole transition matrix element for the exciton.
In terms of the $|D\rangle$, $|X\rangle$, $|Y\rangle$, and $|Z\rangle$ exciton basis, we can express the exciton transition dipole matrix element as,
 \be 
\bm P_{j}(\bm K) = \sum_{i=x,y,z}[C_{i}^j(\bm K)]^\ast \bm P_i =  P_{cv}  \sum_{i=x,y,z}[C_{i}^j(\bm K)]^\ast  g_i \hat{\bm n}_i 
\label{eqDipolesinDXYZbasis}
\ee
 where $\hat{\bm n}_i$ are the unit vectors along the $x$, $y$ and $z$ directions, 
 $ P_{cv}= -i \langle S|\hat{\bm p}_z|z\rangle$ is the Kane momentum matrix element,
  assumed equal for the three p-like conduction band states, $|x\rangle$, $|y\rangle$ and $|z \rangle,$  while $g_i$ is a dimensionless parameter giving the relative magnitudes of the transition dipole matrix elements of the $|X\rangle$, $|Y\rangle$ and $|Z\rangle$ exciton basis states. 
  In our calculations we assumed that the $X, Y,$ and $Z$ exciton basis states have transition dipoles of equal relative magnitude,  $g_x=g_y=g_z$.  (Expressions  for $g_i$ accounting for crystal field effects within a six-band $\bm K \cdot \bm P$ model   can be found in Ref.\citenum{Sercel2019jcp}).

To find the absorption coefficient  we need to find vector potential of light $\bm A^m$ inside the perovskite. 
We first consider the incident light   to be circularly polarized in air with helicity $\pm 1$, corresponding to angular momentum $\pm 1$ along the $\bm  k_{ph}$ direction.  However, accounting for refraction, the angle of propagation of the light in  the perovskite,
 $\theta_{mat},$ is modified from the exterior angle of incidence $\theta$.  Neglecting linear birefringence effects for simplicity, this is given    by Snell's law,
$\sin\theta = n_{\rm mat} \sin\theta_{mat}$.
  Moreover the amplitude of the transmitted electric field  or vector potential is modified by the polarization-dependent Fresnel amplitude transmission coefficients $ t_{\parallel}$ and $t_{\perp}$ for  field components parallel and perpendicular to the plane of incidence, respectively, which are shown in Figure S3 in the Supporting Information.
 
\begin{figure}[h!]
\centering
\includegraphics[width=14cm]{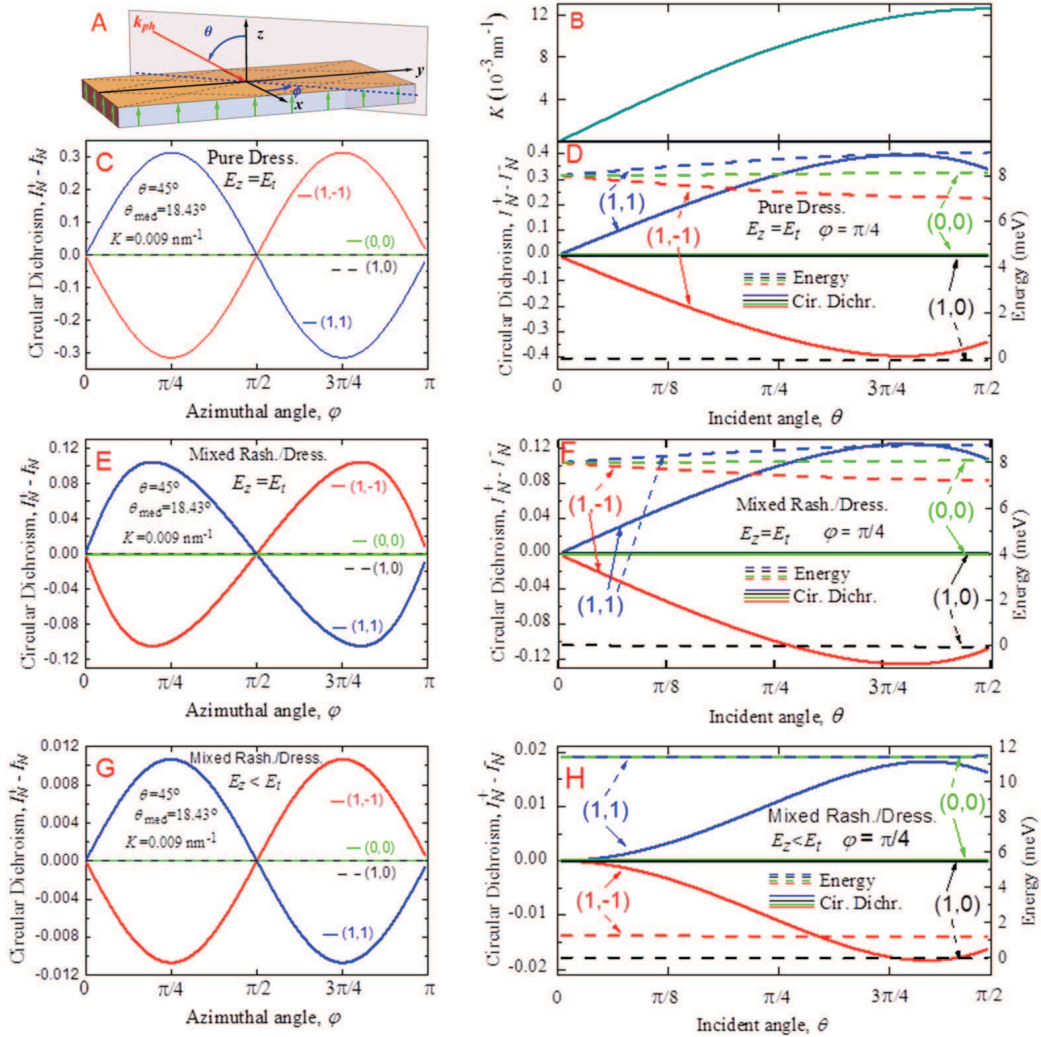}
\caption{
CD signal versus light incidence geometry in a 2D perovskite 
with degenerate triplet states at $\bm K = 0$.  The schematic in  A shows circularly polarized  light   incident at polar angle $\theta$ measured from the layer normal, at azimuth angle $\phi$ measured from the $x$ axis.   The effective masses are $m_e=m_h = 0.25\,m_0$ throughout.  In panels C-F, calculations are performed assuming degenerate  triplet levels at $\bm K=0$ at energy $\Delta =8$\,meV
 above the dark state at energy $E_D=0$\,meV as in Fig. \ref{FigTrousers1}.    Panels C and E show the CD for fine structure level versus $\phi$   for a  fixed polar angle $\theta=45^\circ$,   while panels D and F show the CD versus   $\theta$  for  fixed $\phi=45^\circ$. Panels C, D   reflect pure Dresselhaus spin texture with $\alpha = 0$  and $\beta = 156~{\rm meV}\cdot{\rm nm}$, while  panels E,F are calculated for mixed Rashba/Dresselhaus spin textures with parameters  with $\alpha = 117~{\rm meV}\cdot{\rm nm}$  and $\beta =39~{\rm meV}\cdot{\rm nm}$. Panels G-H are calculated as in E,F, but with crystal field splitting of the bright triplet, reflected in $E_t=11.4$\,meV, $E_z=1.2$\,meV.
}
\label{Fig4CD}
\end{figure}

   At large incidence angles $\theta$ the degree of circular polarization of the transmitted light is reduced because the field component perpendicular to the plane of incidence has lower transmission that the field parallel  component.  This effect  needs to be taken into account since the polarization of the incident circular  light is set before refraction into the sample.  
    For circular light with helicity $\hat{e}_\pm$ propagating as shown in the inset of Fig. \ref{Fig4CD}, the  light transmitted into the sample 
    has amplitude (see the Supporting Information section),
 \begin{align}
 \bm A^m_\pm(\theta_{mat},\phi)=  A_0 \left( t_{\parallel} (\theta_{mat}) \hat{e}_{\parallel}(\theta_{mat},\phi) \pm i  t_{\perp}(\theta_{mat}) \hat{e}_{\perp}(\theta_{mat},\phi)\right)~.
 \label{eqLightVecMat}
 \end{align}
 
  Using these expressions we can write the strength of interaction of the exciton state $j$ with the vector potential $\bm A_\pm^m(\theta_{mat},\phi)$, dropping common factors, as,
 \begin{align}
 I^{\pm}(j) =|\bm A_\pm^m(\theta_{mat},\phi)\cdot \bm P_{j}(\bm K)|^2  
 \end{align}
 It is useful to  define a normalized interaction strength for a given set of angles. This should be normalized  by  the square of the magnitude of the light field, $|A_\pm^m(\theta_{mat},\phi)|^2$.  As discussed in the Supporting information section, we normalize the magnitude of the transition dipole of each exciton state to the average norm-squared dipoles of the  $X,Y,Z$ basis states, which we denote as  $\tilde{f}_N |P_{cv}|^2$, where
$ \tilde{f}_N = (g_x^2+g_y^2+g_z^2)/3$.  
 Then the normalized light-matter interaction strength for exciton state $\bm K$, $j$ is,
 \begin{align}
 I^{\pm}_N(j) =\frac{|\bm A_\pm^m(\theta_{mat},\phi)\cdot \bm P_{j}(\bm K)|^2  }{|A_\pm^m(\theta_{mat},\phi)|^2  \tilde{f}_N |P_{cv}|^2}
 \end{align}
 With this definition we define the magnitude of circular dichroism for the state $j$ as,
\begin{align}
{\cal P}_j = I_{N}^{+}(j) - I_{N}^{-}(j)
\label{eqCDdefined}
\end{align}
We note that this definition coincides with the degree of polarization for the exciton state in a cubic perovskite  with total angular momentum $F_z = \pm 1$ interacting with circularly polarized light propagating in the $+\hat{z}$ direction.

We consider the sample, depicted in  Fig. \ref{Fig4CD} A,  to be oriented in the $x, y$ plane, with the 2D layers parallel to the top surface of the sample which is normal to the $z$ direction.  The sample has    two-fold rotational symmetry about the $z$ axis,  with   inversion symmetry broken along the $+\hat{z}$ direction,   and mirror symmetry through the $x,z$ and  $y,z$ planes.   The sample is assumed to have refractive index $n_{mat}$ and   to be  illuminated from air  at a polar incidence  angle $\theta$ measured from the   vertical $z$   axis, in a plane of incidence defined by azimuthal angle $\phi$ measured from   $x$.
 
 In Figure \ref{Fig4CD}  we  show  plots of the fine structure  energy splitting and the CD signal for various light incidence geometries and sample parameters.  In the figure, the refractive index of the perovskite  is assumed   to be $n=\sqrt{5}$ and the wavelength of light resonant with the exciton transition is taken as $\lambda_0=500$nm  in free space. 
 The schematic in Fig \ref{Fig4CD} panel A shows circularly polarized  light   incident at polar angle $\theta$ measured from the layer normal, at azimuth angle $\phi$ measured from the $\hat{x}$ symmetry axis.  As the polar angle increases, the exciton COM momentum increases,  as shown in panel B.    The exchange constant used in the calculations, taken  as  $w=12$ meV, reflects the measured value in the 2D hybrid organic
 $\rm (C_6H_{13}NH_{3})_2PbI_4$.\cite{Tanaka2005}  In panels C-F, we neglect crystal field splitting resulting in degenerate  triplet exciton levels at $\bm K=0$ at energy $E_t =8$ meV  above the dark state at energy $E_D=0$ meV as in Fig. \ref{FigTrousers1}. Panels C and E show the  CD signal, Eq. \ref{eqCDdefined}, calculated for each fine structure level versus the azimuth angle   for a  fixed polar angle $\theta$ of 45$^\circ$,   while panels D and F show the CD versus the polar angle  for a fixed azimuth angle of 45$^\circ$. The upper
\FloatBarrier
 \noindent  two panels, (C,D)   are calculated  for pure Dresselhaus spin texture with $\alpha = 0$  and $\beta = 156~{\rm meV}\cdot{\rm nm}$ , corresponding to  a Rashba energy for the conduction band    $E_R=40$ meV following  Vardeny, et al., Ref. \citenum{VardenySciAdv}. The  middle panels (E,F) are calculated for mixed Rashba/Dresselhaus spin textures with parameters  with $\alpha = 117~{\rm meV}\cdot{\rm nm}$  and $\beta =39~{\rm meV}\cdot{\rm nm}$ , and show a reduction in the CD relative to the pure Dresselhaus case as expected.  The maximum CD in the pure Dresselhaus case occurs along azimuths at 45 degrees from the mirror planes as expected, while for the mixed case the max CD azimuths are slightly shifted  reflecting  warping of the constant energy  contours for angular momentum textures of mixed character.  
  Panels C,D,E show that the  CD generally increases with polar angle$\theta$ as the in-plane exciton  momentum increases; the fall off as $\theta$ approaches 90 degrees reflects the decreasing degree of circular polarization of the transmitted light described in Eq. \ref{eqLightVecMat} as the polar angle increases.  
 The maximum CD signal is primarily limited by the angle of propagation of light in the material, $\theta_{mat}$, given the assumption that illumination is from the top surface of the 2D layer sample.  Under conditions of side illumination, the  CD signals in Fig. \ref{Fig4CD} would be significantly larger; the maximum CD signal for $\bm k_{ph}$ side incident  along the $ \hat{x} + \hat{y}$ direction for  the material parameters reflected in panel D  would be $\pm 1$  for the $F_{z_B}=\pm 1$ exciton sublevels.  
 
 In panels G-H of Fig \ref{Fig4CD} we repeat the calculations shown in panels E-F, but now take account of  the effect of  crystal field splitting.   Following the measurements of Refs.  \cite{Tanaka2005,KataokaPRB1993} we set the upper bright doublet at $E_x=E_y=11.4$\,meV and the lower bright singlet $E_z=1.2$\,meV above the dark singlet at energy $0$\,meV. Since the CD signal results from 
 mixing of the $Z$ and the $Y^\prime$ exciton states under the action of the effective magnetic field as described above, additional  fine structure splitting between these states causes a reduction in the CD signal by over an order of magnitude.

The analysis so far provides a set of definitions for describing the CD of an individual exciton fine structure level $j$.  In practice, the exciton transitions are spectrally broadened so that the  individual  exciton sublevel transitions may not be individually resolvable. We model the effect of linewidth broadening by convolving the absorption spectra with a Gaussian line-shape function,
\begin{align}
G(E) = \frac{1}{\sqrt{2 \pi \sigma}} e^{-E^2/(2 \sigma^2)}
\end{align}
where the  full-width at half maximum linewidth (LW) is given by $2 \sqrt{2 \ln2}  \sigma$.  We then calculate a normalized absorption spectrum for plus and minus circularly polarized incident light  as a function of energy,  $I^{\pm}_N(E)$, parametrically as a function of the angles $\theta$ and $\phi$: 
\begin{align}
I^{\pm}_N(E) = \frac{1}{I_{max}} \sum_j  I^{\pm}_N(j) G(E-E_j)~.
\label{eqAbslineshape}
\end{align}
In this expression, the sum is taken over the four exciton sub-levels $j$  for a given $\bm K$, and  a normalization factor $I_{max}$ is included to normalize the  peak absorption to unity.  Then the spectral CD signal is calculated as the difference of the normalized absorption spectra for positive and negative circularly polarized light:
\begin{align}
P(E) = I^{+}_N(E) -I^{-}_N(E) ~.
\label{eqAbslineshape}
\end{align}
It should be noted that the normalized spectral CD thus defined has a derivative line-shape  similar to  what is observed in the phenomenon of magnetic CD. \cite{Kuno1998MCD}   Using this definition, which corresponds to the anisotropy factor defined in Ref. \citenum{SargentNatureReview20}, the ellipticity per unit absorbance can be determined at the exciton resonance as $\Theta = P(E) \ln10/4 (180^\circ/\pi)$.\cite{SargentNatureReview20}
 It is straightforward to show that in the limit that the level splitting between the $F_{z_B}=\pm1$ transitions is small compared to the LW, the maximum spectral CD signal occurs at energy $\pm \sigma$ above and below the centroid  $E_c$ of the two transitions. 
 
\begin{figure}[h!]
\centering
\includegraphics[width=15cm]{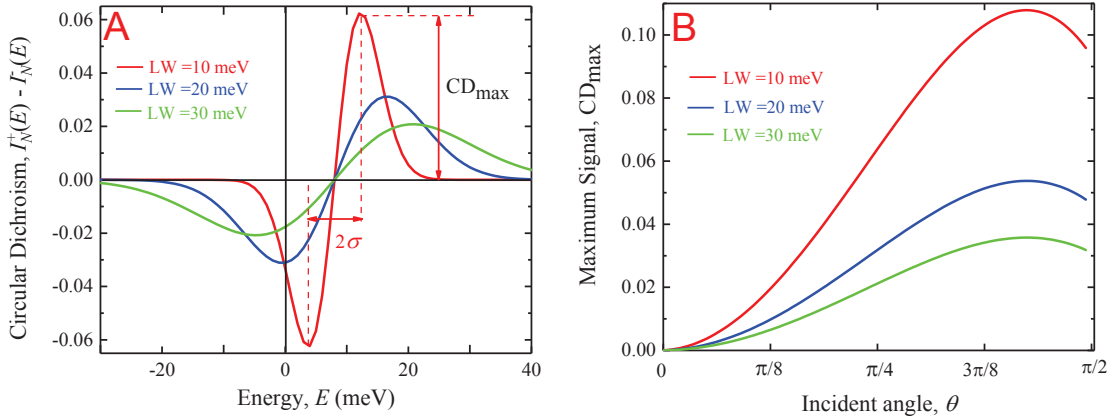}
\caption{Spectral CD signal in 2D  perovskite.  Calculations represent the material parameters of Figure \ref{Fig4CD}(C,D).
 Panel A shows the CD spectra $P(E) = I_+(E) - I_-(E)$ for   fixed azimuth $\phi=45^o$ and     polar angle $\theta=45^o$   as defined in the schematic in Fig. \ref{Fig4CD}A.  Spectra are calculated assuming a Gaussian line-shape for three values of the full width at half maximum LW (LW = $2 \sqrt{2 \ln2} \sigma$)  and are normalized to give a peak absorption $I^\pm$ of unity as described in the text. 
 Transition energy is plotted relative to the energy of the dark exciton at $\bm K=0$.   The zero crossing occurs at the average energy of the $F_{z_B}=\pm1$ states.  The maximum  $\rm|CD|$  occurs at $\pm \sigma$  from the zero crossing and is marked for the LW=10\,meV curve  in the figure.
 The value of the maximum CD on the higher energy side is shown  in panel B as a function of  polar angle $\theta$ for fixed azimuth angle $\phi=45^o$.
}
\label{Fig5spectralCD}
\end{figure}

  Figure \ref{Fig5spectralCD} shows spectral CD results calculated using the  material parameters of Fig \ref{Fig4CD}(C,D), which reflect a degenerate bright triplet finestructure at $\bm K=0$. Panel (A) shows CD spectra calculated for   fixed azimuth and     polar angles of  45 degrees as defined in the schematic in Fig. \ref{Fig4CD} (A),  for several values of the LW=10, 20, 30\, meV.  The spectra exhibit the expected derivitive line shape function centered about the $\bm K=0$ energies of the $F_{z_B}=\pm 1$ states. 
   Panel (B) shows the  maximum CD  versus  polar angle $\theta$  for fixed azimuth angle   $\phi =45^o$.
  
\FloatBarrier

\section{Summary and conclusions } 
We have shown that
perovskite layers can exhibit   CD which is observable in the off-normal optical excitation configuration  and which does not require either an external magnetic field or chiral molecules on the layer surface.  Polarization dependent absorption is expected to occur due to the combined effect of  Rashba splitting, in-plane crystal symmetry breaking,  and the effect of the exciton momentum  on its fine structure.  The fine structure    can be understood in terms of Zeeman splitting due to a $\bm K$-dependent  effective magnetic field, which can lead to splitting of   helical  excitonic states  in systems of orthorhombic or lower crystal symmetry that exhibit Rashba spin splitting effects. 
The selective  coupling of these helical exciton states  to helical light is responsible for the c circular dichroism. 

 Using   available measured  values for the exchange and Rashba parameters  determined experimentally in lead-iodide based 2D HOIS systems \cite{VardenySciAdv,VardenyNatComm2020,Tanaka2005,KataokaPRB1993}, we showed that CD on the order of 10\% for spectrally resolved fine structure transitions  can occur  for top illumination conditions  in orthorhombic perovskites with inversion symmetry breaking normal to the 2D layers.  For optimal side illumination conditions, spectrally resolved exciton fine structure levels may exhibit CD approaching 100\%.
 
 Several effects have been shown to reduce the degree of observable CD.
The principle  limitation  of the 
  CD signal magnitude is the out-of-plane propagation of light given the assumption that the sample is illuminated   from the top surface. Refractive steering of the light 
reduces the  CD signal relative to optimized side illumination conditions by about an order of magnitude.  
 Second, while the largest CD signal  occurs for a pure 2D Dresselhaus-like angular momentum texture,   
a mixed 2D Rashba and 2D Dresselhaus character reduces  its magnitude. 
Additionally, the CD is shown to be reduced by crystal field splitting of the bright triplet exciton levels.   
due to the decrease in their mixing  as the triplet degeneracy  is already broken.   
 Finally,  in the case that the fine structure transitions are not individually spectrally resolved, the observable CD is reduced by line broadening effects.
 
 Our calculations demonstrate that even with spectral line broadening effects  at the level  of tens of meV,  CD   should  be readily observable in   difference spectra which take the form of a  line-shape  derivative function, familiar from magnetic CD studies. Moreover, under optimized side illumination conditions, the  CD is expected to be larger by an order of magnitude relative to the  CD observable under top illumination conditions.   Our results  demonstrate the intriguing possibility of observing circular dichroism in non-chiral perovskite structures such   layered-2D perovskites and nanoplatelets.

\bigskip
\textbf{ACKNOWLEDGEMENTS}
P.C.S. and Z.V.V. acknowledge support from the Center for Hybrid Organic Inorganic Semiconductors for Energy (CHOISE) an Energy Frontier Research Center funded by the Office of Basic Energy Sciences, Office of Science within the US Department of Energy through contract number  DE-AC36-08G028308.
 Al.L.E. acknowledges support  from the US Office of Naval Research  and the Laboratory-University Collaboration Initiative (LUCI)  program of the DoD Basic Research Office.  The authors acknowledge  Dr. Haoliang Liu for useful discussions,  Aditi Chandrashekar for assistance with manuscript preparation and Prof. D.J. Norris for a critical reading of the manuscript and valuable comments. 

The authors declare no competing financial interests.

{\bf Electronic Supplementary Material Available:}  Diagonalization of the exciton Hamiltonian and  details of the CD calculation including analysis of polarization vectors.


\end{document}